\documentstyle[aps,prl,preprint,floats,epsfig]{revtex}

\long\def\simplex#1#2#3{
\begin{figure}[htbp]
   \begin{center}
   \quad\\[-2.5cm]
   \quad\vskip  2.0 cm
   \hbox{
   \quad 
   \parbox[t]{6.5cm}{ \psfig{figure=#1,width=8.5cm}
   \caption[c]{\small  \label{fig:#2} #3 } }
   }
   \quad
   \end{center} 
\end{figure}
}

\def\BR{{\cal B}}
\def\lkh{{\cal L}}
\def\ss{{\it S}}
\def\bb{{\it B}}

\begin{document}

\preprint{\tighten\vbox{\hbox{\hfil CLEO CONF 99-11}
}}

\title{\Large \bf Update of the search for the neutrinoless decay 
$\tau\to\mu\gamma$}  

% Your author list ***DOES NOT*** go here!
% is goes below where you are instructed to insert it...
\author{CLEO Collaboration}
\date{\today}

\maketitle
\tighten

\begin{abstract} 

We present an update of the search 
for the lepton family number violating  decay 
$\tau \to \mu\gamma$  
using a complete CLEO II data sample 
of $12.6$ million $\tau^+\tau^-$ pairs. 
No evidence of a signal has been found and 
the corresponding upper limit is 
$ \BR( \tau \to \mu\gamma ) < 1.0 \times 10^{-6} $ at 90\% CL,
significantly smaller than previous limits. 
All quoted results are preliminary.

\end{abstract}

\newpage

{
\renewcommand{\thefootnote}{\fnsymbol{footnote}}

% Insert author and address list here
\begin{center}
A.~Anastassov,$^{1}$ J.~E.~Duboscq,$^{1}$ K.~K.~Gan,$^{1}$
C.~Gwon,$^{1}$ T.~Hart,$^{1}$ K.~Honscheid,$^{1}$ H.~Kagan,$^{1}$
R.~Kass,$^{1}$ J.~Lorenc,$^{1}$ H.~Schwarthoff,$^{1}$
% ...
\end{center}
 
\small
\begin{center}
$^{1}${Ohio State University, Columbus, Ohio 43210}\\
% ...
\end{center}
 
\setcounter{footnote}{0}
}
\newpage

This note summarizes results of the final search  
of the complete CLEO II
data sample for the lepton number violating decay
$\tau\to\mu\gamma$. Although there are many possible $\tau$ decay channels 
which do not conserve the leptonic quantum number, the decay 
$\tau\to\mu\gamma$ is 
favored by most theoretical extensions of the Standard 
Model~\cite{Stroynowski}. 
The most optimistic predictions for rates
of such decays are based on
the supersymmetric models~\cite{Barbieri,Hisano} and on the left-right 
supersymmetric models\cite{Mohapatra}.
In a recent calculation~\cite{Hisano2}, based on the
Minimal Supersymmetric Standard Model with right-handed neutrinos,
a value of $2\times 10^{-6}$ for the
branching fraction for the decay $\tau\to\mu\gamma$ 
has been obtained for some ranges of model parameters.
In general, the expectations for 
all other lepton number or lepton flavor violating decays of the $\tau$ 
are at least an order of magnitude lower.  
Experimental searches for the $\tau\to\mu\gamma$ decay 
are limited by statistics, i.e., the number of observed $\tau$ decays.
The smallest upper limit~\cite{CLEO2} of 
$\BR(\tau\to\mu\gamma) < 3.0\times 10^{-6}$ at 90\% CL
has been published by the CLEO Collaboration
using $4.24$ million $\tau^+\tau^-$ pairs.

In this analysis we use a data sample from the reaction 
$e^+e^-\to\tau^+\tau^-$  
collected at CESR at or near the energy of the
$\Upsilon(4S)$. The data correspond to a total 
integrated luminosity of 
$13.8\ {\mbox{fb}}^{-1}$ 
and contain about 12.6 million $\tau^+\tau^-$ 
pairs. The event selection follows the 
procedure used in the previous search~\cite{CLEO2}.
We study events with a 1-vs-1 topology, where the 
signal candidate $\tau$ decays into $\mu\gamma$
and the tag side includes all standard $\tau$ decays 
into one charged particle, any number 
of photons and at least one neutrino. 

We select $\tau^+ \tau^-$ pair events with exactly 
two good charged tracks,  
with total charge equal to zero, and with 
an angle between the charged tracks greater than $90^{\circ}$. 
Because radiative $\mu$-pair production 
provides high background rates, we allow
only one identified muon per event.  
In addition, each candidate event
must have exactly one photon
separated by more than $20^{\circ}$ from the closest charged track 
in the lepton hemisphere.
This photon must lie in the calorimeter barrel 
(i.e., $|\cos\theta_\gamma|<0.71$,
where $\theta_\gamma$ is an angle between the photon and beam directions),
have a photon-like lateral profile 
and have energy deposition in the calorimeter greater than 300 MeV. 
This minimum energy cut is 
dictated by the kinematics of a 2-body $\tau$ decay. 
The angle between the direction of the photon and the momentum of 
the muon track must satisfy 
$0.4<\cos\theta_{\mu\gamma}<0.8$, where the upper limit is 
again dictated by kinematics, and the lower limit is obtained by 
optimizing the signal-to-background ratio.

The main sources of background in the selected samples are due 
to $\mu$-pair production, radiative $\tau\to\mu\gamma\nu\nu$ 
decays, and two-photon processes. To minimize these backgrounds,
we require that the cosine of the angle 
between the total missing momentum of the event and the momentum 
of the tagging particle be greater than 0.4. 
The missing momentum is calculated as the negative 
of the sum of momenta of the
two charged tracks and all showers detected in the 
calorimeter with energies above 30 MeV.
Because there must be at least one  
undetected neutrino on the tag side, the missing momentum in an
event having $\tau\to\mu\gamma$ is expected to fall into the 
tagging track hemisphere, while for all radiative processes the missing
momentum should be uncorrelated with the 
charged track on the tag side.
The neutrino emission on the tag side should also result in a 
large total transverse momentum with respect to the beam direction. 
Thus, to suppress background produced by 
copious two-photon and radiative QED processes, we require that
the total transverse momentum of the event be greater 
than $300\ \mbox{MeV}/c$.
The selection efficiency of all the previous requirements is estimated
as 16.2\%.

Final signal selection criteria are based on kinematic 
constraints since
a neutrinoless $\tau$ decay should have a total energy and an effective 
mass of the $\mu\gamma$ 
consistent with the beam energy and $\tau$ mass, respectively. 
To determine these final criteria, we employ two 
different techniques. First, we follow the method outlined in 
CLEO's previous search~\cite{CLEO2} for the decay $\tau\to\mu\gamma$.
Then we perform a more sensitive analysis based on an unbinned extended
maximum likelihood (EML) fit to the data. 

Following the method described in detail in Ref.~\cite{CLEO2}, 
we parameterize the signal Monte Carlo mass
and energy distributions separately as tailed Gaussian densities.
The energy density is given by:
\begin{eqnarray}
\label{eq:cbl}
f(E) = \left\{
\begin{array}{rr}
\left\{ 
l/ \left[ \eta (-\tilde{E}+l/\eta-\eta) \right]
\right\}^l
\exp(-\eta^2/2)\ ; & \tilde{E}<-\eta\ ; \\
\exp(-\tilde{E}^2/2)\ ; & \tilde{E}>-\eta\ ;
\end{array}
\right. \\
\tilde{E} = \frac{E-E_{beam}}{\sigma_E}\ , \nonumber
\end{eqnarray}
where $\sigma_E$, $\eta$, and $l$ are the fit parameters.
A similar formula for the mass density is obtained by
substituting the symbol ``$m$'' instead of ``$E$'' and
``$m_\tau$'' instead of ``$E_{beam}$'' in the equation above.
The signal region is then defined to be within 
$\pm 3$ standard deviations of
the fitted Gaussian component of the distribution. 
The $\tau$ mass, $m_\tau$,
is taken to be $1.777\ \mbox{GeV}/c^2$, and the beam
energy $E_{beam}$ varies from 5.26 to 5.29~GeV.
To estimate the amount of background expected in the signal region,
we extrapolate the data from the sideband. 
We assume that the background distributions are linear  in 
the vicinity of $m_\tau$ and $E_{beam}$ and define the sideband regions 
to be between 5 and 8 standard deviations as shown in Fig.~\ref{fig:e_vs_m}.
To estimate the background uncertainty associated with this
technique, we vary the sideband definition. The total expected
background in the signal region is estimated as $5.5\pm 0.5$ events.

\simplex{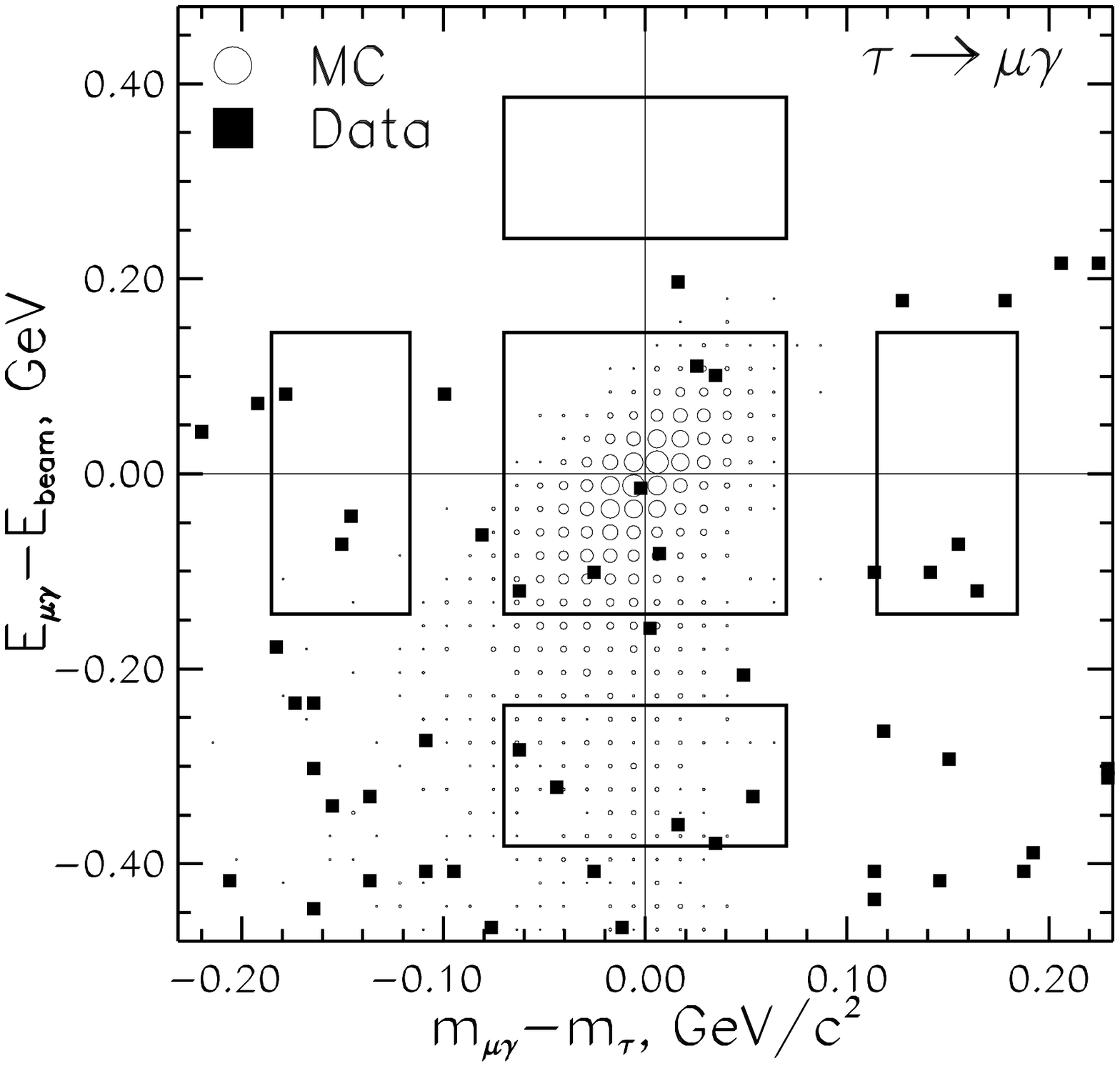}{e_vs_m}
{$(E_{\mu\gamma}-E_{beam})$ vs $(m_{\mu\gamma}-m_{\tau})$ distribution.
Solid squares represent the data, open circles
represent the signal Monte Carlo distribution.}

The upper limit on the $\tau\to\mu\gamma$ branching fraction is 
estimated following the Bayesian prescription~\cite{helene,PDG96}:
\begin{equation}
\label{eq:bayes}
\frac{ e^{-(s_0+b)} \sum_{n=0}^{n_0} \frac{(s_0+b)^n}{n!}}
{ e^{-b} \sum_{n=0}^{n_0} \frac{b^n}{n!} } = 0.1\ ,
\end{equation}
where $s_0$ is an upper limit on the number of events in the signal
region at 90\% CL, $b$ is the expected background rate, and $n_0$ is the
number of observed events. The upper limit on the branching fraction
is then:
\begin{equation}
\label{eq:norm}
\BR(\tau\to\mu\gamma) < \frac{s_0}{2\epsilon N_{\tau\tau}}\ 
\mbox{at 90\% CL}\ ,
\end{equation}
where $\epsilon$ is the event selection efficiency and $N_{\tau\tau}$ 
is the total number of $\tau$-pairs produced. Applying this technique,
we obtain an upper limit on the branching fraction 
$\BR(\tau\to\mu\gamma)$ of $1.8\times 10^{-6}$ at 90\% CL.

The systematic uncertainty in detector sensitivity 
$2\epsilon N_{\tau\tau}$ 
is conservatively estimated as 10\%. 
This uncertainty is obtained by adding
in quadrature uncertainties in
track reconstruction efficiency (3\%), photon reconstruction efficiency (5\%),
cut selection (5\%), luminosity (1.4\%), lepton identification (4\%),
Monte Carlo statistics (1.5\%) and trigger efficiency (5\%).
The upper limit for the branching fraction is also affected by
the uncertainty in background estimate: $\sigma_B=0.5$. 
To incorporate systematic uncertainty
into the upper limit, we assume that the errors related to
$2\epsilon N_{\tau\tau}$ and to the background estimate have Gaussian
distributions and apply a technique described in Refs.~\cite{CLEO2,Cousins}.
This technique reweights the probability~(\ref{eq:bayes})
by a Gaussian probability density of the detector sensitivity
$2\epsilon N_{\tau\tau}$ and a Gaussian probability density of the number 
of background events $b$. The incorporation of these
systematic uncertainties 
increases the upper limit by 1.9\% of itself.

A more sensitive upper limit is obtained by performing 
an unbinned EML fit which takes into account 
the details of the distributions and 
correlations between the mass and
energy of signal event candidates.
The likelihood function is defined as:
\begin{equation}
\label{eq:lkh}
\lkh(s,b) = \frac{e^{-(s+b)}}{N!} \prod_{i=1}^N (s\ss_i+b\bb_i)\ ,
\end{equation}
where $N$ is the number of events in the signal region and its vicinity,
$s$ and $b$ are the numbers of signal and background
events, respectively, and $\ss_i$ and $\bb_i$ are the signal and
background densities, respectively. 
The signal distribution is described by a two-dimensional Gaussian 
and a non-Gaussian tail in energy produced by initial and final 
state radiation. This tail covers the region below the beam energy 
and is modeled by a gamma-function.
\newpage
\begin{eqnarray}
\label{eq:signal}
\ss_i (m,E) & = & 
\frac{A_G}{2\pi\sigma_m\sigma_E\sqrt{1-\rho^2}} \times \nonumber \\
& & \times \exp
\left\{
-\frac{1}{2(1-\rho^2)}
\left[ 
\left( \frac{m-m_\tau}{\sigma_m} \right)^2
- 2\rho 
\left(\frac{m-m_\tau}{\sigma_m}\right)
\left(\frac{E-E_{beam}}{\sigma_E}\right)
+ \left( \frac{E-E_{beam}}{\sigma_E} \right)^2
\right]
\right\} + \nonumber \\
& & + A_T\zeta(m,E)\ ;\\
& & \nonumber \\
\zeta(m,E) & = & \left\{
\begin{array}{cl}
\frac{1}{\sqrt{2\pi}\sigma_m}
\exp\left[ -\frac{1}{2} \left( \frac{m-m_\tau}{\sigma_m} \right)^2 \right]
\frac{1}{\sigma_E\Gamma(\alpha)\beta^\alpha}
\left( \frac{E_{beam}-E}{\sigma_E} \right)^{\alpha-1}
\exp\left[ -\frac{E_{beam}-E}{\beta\sigma_E} \right] & 
\mbox{if $E<E_{beam}$} \\
0 & \mbox{otherwise}
\end{array}
\right. \nonumber
\end{eqnarray}
where $A_G$ and $A_T$ are the relative contributions of the Gaussian
component and the non-Gaussian tail with the sum of $A_G+A_T$ constrained
to unity, $\sigma_m$ and $\sigma_E$ are mass and energy resolutions, 
respectively, $\rho$ is the correlation coefficient,
and $\alpha$ and $\beta$ 
define the shape of the non-Gaussian tail $\zeta(m,E)$.
To obtain the parameters of the signal density $\ss_i$,
we fit the signal Monte Carlo distribution.
The background is parameterized by a function linear in energy with 
the coefficients $a_0$ and $a_1$ obtained from a fit to the data:
\begin{equation}
\label{eq:bckgr}
\bb_i (m,E) = \frac{1}{m_2-m_1} 
\frac{1}{ (a_0-a_1E_{beam})(E_2-E_1) + 0.5a_1(E_2^2-E_1^2) }
\left[ a_0 + a_1(E-E_{beam}) \right]\ ,
\end{equation}
where $(m_1,m_2)$
and $(E_1,E_2)$ are the limits defining the fit region. 
The region within 4 standard deviations
near the beam energy $E_{beam}$
is excluded from the fit to avoid bias caused by the 
possible presence of real signal events in this region.
Uncertainties of the background shape parameters $a_0$ and $a_1$
are estimated by varying the number of bins in the fit region.

The EML fit to the data gives the number of candidates for 
the decay $\tau\to\mu\gamma$ as 1.8 events with an estimated statistical
significance of the signal $1.0$ standard deviations. The fit region, 
shown in Fig.~\ref{fig:e_vs_m}, is defined to be within 10 standard
deviations near the $\tau$ mass and beam energy.
The total number of events in the fit region is 53.

To estimate the upper limit, we use a method~\cite{narsky}
developed for unbinned EML fits.~\footnote{ 
This method assumes a confidence interval to be from 0 to $s_0$
and thus gives a different upper limit than that obtained
by the method of Ref.~\cite{Feldman}. The prescription~\cite{Feldman}
has been developed for problems with integer numbers of observed
signal candidate events and, in its present shape, is inapplicable
to EML fits.}
For every assumed expected number 
of signal events $s$, we generate 10,000 Monte Carlo samples.
For every sample, we generate numbers of signal and background events 
using Poisson distributions and then we generate
positions of these events on the energy-vs-mass plane
using the densities from Eqns.~(\ref{eq:signal}) and (\ref{eq:bckgr}).
For each sample we then perform
an unbinned EML fit to extract the number of signal events, following
the same procedure as for the
data. The confidence level corresponding to
this value of $s$ is defined as a fraction of samples where the
extracted number of events exceeds that observed in the data, 
i.e., 1.8. We repeat this procedure until we find a value of $s=s_0$
that gives a 90\% CL. This value has to be divided by the selection
efficiency and the number of produced $\tau$-pairs in accordance with
Eqn.~(\ref{eq:norm}). The obtained upper limit on the branching
fraction $\BR(\tau\to\mu\gamma)$ is $1.0\times 10^{-6}$ at 90\% CL.

To incorporate systematic uncertainty
into this result, we smear the background shape parameters $a_0$ and $a_1$
within the estimated uncertainties assuming Gaussian distributions
and taking into account the correlation between these two parameters.
We then repeat the procedure described in the previous paragraph, 
the only modification is that now we integrate the likelihood function
over the parameter space of $a_0$ and $a_1$. Because the parameters
of the signal density are known with better accuracy and because
we do not observe a significant signal contribution, the result is
insignificantly affected by uncertainties in the parameters of the
signal density function. In addition to smearing the parameters, 
we integrate the quantity
$1/(2\epsilon N_{\tau\tau})$ assuming a Gaussian
distribution for the detector sensitivity $2\epsilon N_{\tau\tau}$
with a relative standard deviation equal to the estimated
systematic uncertainty, i.e., 10\%. 
A conservatively estimated relative effect of the 
systematic uncertainties on the upper limit is below 10\%. 
Since this computation is not
yet finished, we quote a result without inclusion of systematic 
uncertainties.

The selection efficiencies, numbers of events, and upper limits
calculated 
for both techniques are given in Table~\ref{tab:results}.
This result is limited by the total integrated luminosity and
represents a significant improvement over the previous 
analysis~\cite{CLEO2}. It also, for the first time, restricts the parameter
space of some versions of the MSSM models~\cite{Hisano2}.

\begin{table}[bthp]
\caption{\label{tab:results}}
{Selection efficiencies, numbers of events, and upper limits.
%calculated with and without systematic errors.
}
\begin{center}
\begin{tabular}{|lcc|}\hline
Method 		& Method of Ref.~\cite{CLEO2} & Unbinned EML fit \\ \hline
MC efficiency, $\epsilon$	& 12.7\%	& 15.2\% \\
Number of signal events		& $n_0=6$	& $s=1.8$ \\
Expected background rate, $b$	& $5.5\pm 0.5$	& -        \\
Statistical significance of the signal	
				& -		& $1.0\sigma$ \\
Upper limit at 90\% CL, $s_0$	& 5.8		& 3.8 \\
Upper limit for $\BR(\tau\to\mu\gamma)$ at 90\% CL
				& $1.8\times 10^{-6}$ 
				& $1.0\times 10^{-6}$ \\
\hline	
\end{tabular}
\end{center}
\end{table}

We gratefully acknowledge the effort of the CESR staff in providing us with
excellent luminosity and running conditions.
J.R. Patterson and I.P.J. Shipsey thank the NYI program of the NSF, 
M. Selen thanks the PFF program of the NSF, 
M. Selen and H. Yamamoto thank the OJI program of DOE, 
J.R. Patterson, K. Honscheid, M. Selen and V. Sharma 
thank the A.P. Sloan Foundation, 
M. Selen and V. Sharma thank the Research Corporation, 
F. Blanc thanks the Swiss National Science Foundation, 
and H. Schwarthoff and E. von Toerne thank 
the Alexander von Humboldt Stiftung for support.  
This work was supported by the National Science Foundation, the
U.S. Department of Energy, and the Natural Sciences and Engineering Research 
Council of Canada.

\end{document}